\documentclass[journal=langd5,manuscript=article]{achemso}

\usepackage[version=4]{mhchem}
\usepackage{gensymb}
\usepackage{times,mathptmx}
\usepackage{sectsty}
\usepackage{balance}
\usepackage{float}
\usepackage{color,graphicx,amssymb,stmaryrd,upgreek}
\usepackage[dvipsnames]{xcolor}
\usepackage{graphicx} 
\graphicspath{{./Paper_Revised/}}
\usepackage{natbib}

\DeclareUnicodeCharacter{0301}{\'{e}}

\title{Characterization and Optimization of Fluorescent Organosilica Colloids for 3D Confocal Microscopy Prepared Under 'Zero-Flow'}

\author{Ruth A. Crothers}
\affiliation{Institute for Molecules and Materials, Radboud University Heyendaalseweg 135, 6525 AJ Nijmegen, The Netherlands}
\altaffiliation{Department of Chemistry, Physical and Theoretical Chemistry Laboratory, University of Oxford, South Parks Road, Oxford OX1 3QZ, United Kingdom}
\author{Nicholas H. P. Orr}
\affiliation{Laboratoire Charles Coulomb UMR 5221, Universit\'{e} de Montpellier, F-34095 Montpellier, ́France}
\altaffiliation{Department of Chemistry, Physical and Theoretical Chemistry Laboratory, University of Oxford, South Parks Road, Oxford OX1 3QZ, United Kingdom}
\author{Berend van der Meer}
\altaffiliation{Department of Chemistry, Physical and Theoretical Chemistry Laboratory, University of Oxford, South Parks Road, Oxford OX1 3QZ, United Kingdom}
\author{Roel P. A. Dullens}
\affiliation{Institute for Molecules and Materials, Radboud University Heyendaalseweg 135, 6525 AJ Nijmegen, The Netherlands}
\altaffiliation{Department of Chemistry, Physical and Theoretical Chemistry Laboratory, University of Oxford, South Parks Road, Oxford OX1 3QZ, United Kingdom}
\author{Taiki Yanagishima}
\affiliation{Department of Physics, Graduate School of Science, Kyoto University, Kyoto, 606-8224, Japan}
\altaffiliation{Department of Chemistry, Physical and Theoretical Chemistry Laboratory, University of Oxford, South Parks Road, Oxford OX1 3QZ, United Kingdom}
\email{yanagishima.taiki.8y@kyoto-u.ac.jp}

\begin{document}

\abstract{We optimize and characterize the preparation of 3-trimethoxysilyl propylmethacrylate (TPM) colloidal suspensions for three-dimensional confocal microscopy. We revisit a simple synthesis of TPM microspheres by nucleation of droplets from pre-hydrolyzed TPM oil in a 'zero-flow' regime, and demonstrate how precise and reproducible control of particle size may be achieved via single-step nucleation with a focus on how the reagents are mixed. We also revamp the conventional dyeing method for TPM particles to achieve uniform transfer of a fluorophore to the organosilica droplets, improving particle identification. Finally, we illustrate how a ternary mixture of tetralin, trichloroethylene and tetrachloroethylene may be used as a suspension medium which matches the refractive index of these particles while allowing independent control of the density mismatch between particle and solvent.}

\section{Introduction}
Colloidal particles have been widely used as a model system to study condensed matter phenomena, including crystallization \cite{Lekkerkerker2005,Palberg2014,Chen2021}, gelation \cite{Trappe2004,Buzzaccaro2007,Lu2008,Tsurusawa2019} and glass formation \cite{Pusey1991,Simeonova2006,Hunter2012,Hallett2018}. 
The development of confocal microscopy has allowed direct observation of fluorescently labeled colloidal systems in three dimensions; with the right choice of particle density and a solvent with a matching refractive index, the positions of all particles in an observation window may be tracked over time with single-particle resolution \cite{Kegel2000,Prasad2007}, offering the prospect of 'complete' knowledge of the structural state of the system.

But the most common colloidal systems e.g. poly-methyl methacrylate (PMMA) \cite{Bosma2002}, silica \cite{Blaaderen1995}, poly(n-isopropylacrylamide) (PNIPAM) \cite{Yunker2014}, polystyrene \cite{Rutgers1996,Lu2013} have distinct hurdles to their use in three-dimensional studies. These range from inherent softness in refractive index matching solvents (e.g.PNIPAM), to the expertise required to make the particles from scratch (e.g. PMMA), and difficulty in matching refractive index and/or density with a solvent (e.g. silica, polystyrene). As a more accessible solution, Sacanna {\it et al.} \cite{Sacanna2010,Sacanna2011}, and later by an alternative pathway Van der Wel {\it et al.} \cite{VanDerWel2017} proposed the condensation of 3-(trimethoxysilyl)propyl methacrylate (TPM) in a base to produce monodisperse colloidal spheres at the micron-scale. Further work by Liu {\it et al.} demonstrated that common organic solvents and a solvent transfer procedure could be used to suspend these TPM particles and create a refractive index matched suspension suitable for confocal microscopy \cite{Liu2016,Liu2019}. TPM as a colloidal material has been repeatedly shown to be versatile in the creation of different particle shapes \cite{Sacanna2010}, tunable interactions \cite{Opdam2020} and fluorescence profiles \cite{Liu2019,Yanagishima2021}, making it an ideal system for 3D microscopy.

Yet, there are issues that remain to be addressed. In brief, synthesis of TPM particles involves the formation of charge-stabilized oil-in-water emulsion via hydrolysis and condensation of TPM; these droplets are then subsequently crosslinked into particles \cite{Sacanna2011}. Monodisperse TPM emulsion droplets were first made by Sacanna {\it et al.} as precursors to dimpled particle formation \cite{Sacanna2010}. In their procedure, they describe the hydrolysis of TPM in water before adding ammonia to nucleate; hydrolyzed TPM is then added for further growth. Van der Wel {\it et al.} \cite{VanDerWel2017} described the particle formation in a fuller manner, only now adding TPM oil to a stirring solution of base for simultaneous hydrolysis and condensation: this is now considered the 'conventional' TPM synthesis. However, certain aspects remain difficult to control \cite{VanDerWel2017, Middleton2019}, particularly in producing particles of an exact target size. Middleton {\it et al.} \cite{Middleton2019} have demonstrated that stirring speed directly impacts final particle size, but in a non-monotonic fashion. The effect of stirring was also noted by Neibloom {\it et al.} \cite{Neibloom2020}, who noted that increased particle collisions may cause particle aggregation. Clearly, the flow field has a strong impact on target diameter and size distribution. But an identical flow profile is difficult to replicate between batches, as it is dictated by not only the speed of rotation of the stirrer but its shape and position, as well as vessel geometry. This is why very similar reaction conditions in different works lead to variations in particle sizes up to a factor of 2 (or a factor of 8 in volume), as reviewed in the Results section. Therefore, it is desirable to revisit Sacanna's method\cite{Sacanna2010}, but data showing the relationship between particle size and pre-hydrolyzed TPM concentration, and how reproducibly specific particle sizes may be obtained in the absence of homogenization, remains unavailable.

There are also issues to be resolved with the fluorescent labeling of particles. Established methods of dyeing TPM particles with a fluorophore dissolved in DMSO \cite{Liu2019} lead to an uneven distribution of dye throughout the batch. This results in a small population of overly bright particles which can saturate the signal in confocal microscopy images, leading to inaccuracy in particle identification.

A further limitation of existing model systems is achieving simultaneous refractive index and density matching. TPM can be refractive index matched with tetralin and trichloroethylene, however the composition that achieves an exact refractive index match between the particle and solvent is not the same as the density matching point. To achieve density matching, the refractive index match must be compromised. This leads to difficulties when studying the dynamics of dense particle suspensions as a function of gravitational length. There are many examples of such studies \cite{Cheng2001,Simeonova2004,Kim2013,Harich2016}; this issue is particularly acute for studies such as that by Wood {\it et al.} \cite{Wood2018} which require control over density mismatch while retaining a high level of tracking accuracy deep within dense samples.

In this article, we modify and optimize these aspects of TPM particle preparation for 3D confocal microscopy, guiding potential users all the way from TPM monomer to fluorescently labeled, index-matched particles. We revisit the protocol of Sacanna {\emph et al.}\cite{Sacanna2010} with some modifications, finding that single-step nucleation of droplets in a 'zero-flow' regime gives precise control over target particle sizes within a certain concentration window, provided reagent mixing is swift and that nucleation occurs under quiescent conditions. We also show how exchange of DMSO for a less viscous solvent improves the distribution of fluorophore throughout a batch of TPM particles. Finally, we introduce a ternary organic solvent mixture for suspending TPM particles which allows a close refractive index match to TPM particles with independent control of density mismatch between particle and solvent.

\section{Experimental Methods}

\subsection{Materials}
3-(trimethoxysilyl)propyl methacrylate (TPM, $\geq$97$\%$, Sigma Aldrich), ammonium hydroxide solution (28$\%$ vol. \ce{NH3} in \ce{H2O}, Alfa Aesar), azobisisobutyronitrile (AIBN, BDH, UK),  Cyanine3 NHS Ester (Lumiprobe), Rhodamine B isothiocyanate (RITC, Sigma Aldrich), 3-aminopropyl trimethoxysilane (APS, Sigma Aldrich), acetone (Sigma Aldrich), trichloroethylene (TCE,  $\geq$99.5$\%$, Sigma Aldrich), 1,2,3,4-tetrahydronaphthalene (TTL,  anhydrous, 99$\%$, Sigma Aldrich) and tetrachloroethylene (PERC, 99$\%$, Alfa Aesar) were used as received. A polyisobutylene succinic anhydride stabilizer, OLOA 11000, was provided by Eric Dufresne and Azelis. Double-distilled water was sourced from a Millipore Direct-Q 3 ultrapure water unit.

\subsection{TPM Droplet Formation}
\begin{figure}[H]
\includegraphics[width=15cm]{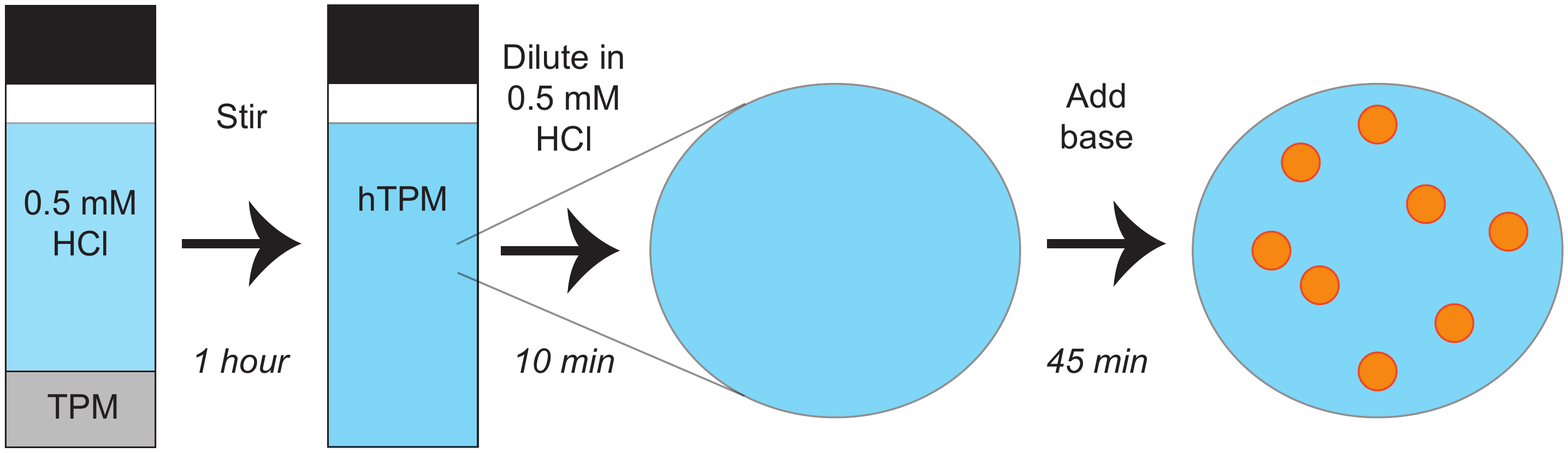}
\centering
\caption{Schematic of droplet synthesis using pre-hydrolyzed TPM in the absence of flow. Note the sample is only stirred to pre-hydrolyze the TPM; particle formation takes place in quiescent conditions.}
\label{fig:schematic}
\end{figure}

We pre-hydrolyze TPM oil by adding 2 ml TPM to 20 ml 0.5mM HCl. The emulsion was left to stir vigorously for 1 hour after which the solution goes transparent, and we considered all TPM monomers to be fully hydrolyzed. We note that hydrolysis in these mildly acidic conditions significantly accelerates the hydrolysis, as widely noted for other silanes \cite{Hermanson2008}, while avoiding any significant effect on subsequent steps at higher pH. For reference, existing methods refer to TPM pre-hydrolysis under neutral conditions for 24 hours \cite{Sacanna2011}. The resulting hydrolyzed TPM solution (hTPM) was further diluted with 0.5mM HCl in a 150 ml round bottom flask to a target concentration and stirred at 600 rpm for 10 min. Exact quantities of reactants are shown in Table \ref{tab:reactants}. Stirring was then stopped and the flask left to settle until contents were stationary. To form TPM droplets, a volume of 0.028\% \ce{NH4OH} was added swiftly, as one volume, to the flask using a 25 ml plastic pipette (Starlab) and filler. Upon addition of the base, the flask was left to stand for 45 minutes to allow droplet formation to take place. A schematic is shown in Figure \ref{fig:schematic}.

\begin{table}
\centering
\begin{tabular}{|c|c|c|c|c|c|c|c|c|}
\hline
0.028$\%$  \ce{NH4OH}/ ml      & 25         & 25         & 25           & 25           &25            &25         &25             &25\\ \hline
0.5mM HCl/ ml           & 24         & 22         & 22           & 21           &20            &19         &18             &17\\ \hline
hTPM/ ml                & 1          & 2          & 3            & 4            &5             &6          &7              &8\\ \hline
[TPM]/ x$10^{-5}$M   & 0.77          & 1.53          & 2.30            & 3.06            &3.83             &4.59          &5.36              &6.12\\ \hline
\end{tabular}
\caption{Quantities of reactants used for initial TPM droplet formation.}
\label{tab:reactants}
\end{table}

In order to reach larger particle diameters, we also briefly document the further addition of hTPM to pre-formed droplets, as performed in previous work \cite{Sacanna2011,VanDerWel2017}. Specifically, we focus on how reproducible the growth path is when a syringe pump is used to add the mildly acidic pre-hydrolyzed TPM solution described above at a set rate. The suspension of pre-formed droplets was stirred at 200 rpm under the assumption that further nucleation of particles in the main part of the particle size population does not occur. The drip rate was kept at 160 $\mu$l/min from a 10 ml syringe mounted on syringe pump (PHD 2000 Harvard Apparatus). This rate was low enough to suppress secondary nucleation. The syringe was connected to a PTFE tube inserted into the reaction vessel through a stopper to minimize evaporation of the base.

In both single-step droplet formation and growth experiments, the droplets were crosslinked with AIBN, adding approximately 1 mg per 5 ml emulsion and stirring for 30 min. It should be emphasized that particle formation has finished at this point i.e. all stirring operations in this protocol are applied {\it after} particle formation has ceased. The suspension was then transferred to an 80 \degree C oven for 3 hours, tumbling every 30 min to prevent coalescence of droplets. The crosslinked batches were washed by repeated centrifugation and decanting of the supernatant.

\subsection{Dye Procedure}
 The procedure for dyeing TPM particles was modified from previous work\cite{Liu2019}; the solvent for dye molecules was changed from DMSO to acetone. The mixture was prepared by combining 2 ml of acetone with 2 mg dye (Cyanine3 NHS Ester or RITC) and 2 $\mu$l APS before stirring in the dark in a closed 14 ml glass vessel for 24 hours. When the droplets are dyed, 2 ml of 5 wt\% F108 was added as described previously to stabilize the droplets\cite{Liu2016}, letting the solution stir for 45 min. We then added 400 $\mu$l of dye solution and the resulting suspension was left to stir for 1 hour before crosslinking.

We also note that uneven dyeing can return if the dye solution was prepared more than a month before use. The linking agent used, APS, reacts with itself over time \cite{VanDerWel2017} causing aggregation of the dye. This can lead to uneven dyeing and brightness when these aggregates are incorporated into particles.

\subsection{Tuning Density Mismatch of Refractive Index Matched Solution}
In previous work, a 60:40 w/w combination of TCE and TTL was used to approximately refractive index and density match crosslinked TPM. PERC has a refractive index, $n_f$, close to that of crosslinked TPM but a much higher density (see Table \ref{tab:properties}). Thus, addition of PERC to a refractive index matched solution of TPM particles allows tuning of the density mismatch between particle and solution without sacrificing the refractive index match. Note that all organic solvents used in the density and refractive index matching of TPM contain 5wt\% OLOA 11000. OLOA 11000 is a charge control agent and stabilizer which has been shown to keep TPM particles stable in non-polar solvents and provide electrostatic screening. A concentration of 5wt\% has been shown to keep the particles hard-sphere-like in their interactions in organic solvents\cite{Yanagishima2021}.

\begin{table}[h]
\centering
\begin{tabular}{|c|c|c|}
\hline
\multicolumn{1}{|l|}{} & \text{$\rho$/g ml$^{-1}$} & \text{$n_f$} \\ \hline
TTL$^a$         & 0.973         & 1.541    \\ \hline
TCE$^a$          & 1.463         & 1.476    \\ \hline
PERC$^a$         & 1.623         & 1.506    \\ \hline
TPM$^b$          & 1.314          & 1.512    \\ \hline
\end{tabular}
 \small
\caption{Refractive index and density of organic solvents used in comparison with crosslinked TPM. $^a$ Data taken from manufacturer website. $^b$ Data taken from\cite{VanDerWel2017}.}
\label{tab:properties}
\end{table}

To demonstrate the use of this solvent mixture to study density-mismatch dependent behavior in dense colloidal systems, we prepared fluorescent TPM particles and added a non-fluorescent TPM surface layer using a method outlined in previous work \cite{Liu2016}. In brief, fluorescent particles are prepared as described above, the only difference being that they are exposed to a 0.5\%wt Pluronic F108 solution before cross-linking. When additional hTPM is added to these particles, the surfactant induces the formation of well-defined lobes on the surface. On cross-linking these lobes, the particles form fluorescent core-shell ‘raspberry’ particles. Subsequent additions of hTPM fill in the space between the lobes. A final cross-linking step results in a spherical core-shell particle with a non-fluorescent shell. This increases the separation between fluorescence profiles of adjacent particles in dense suspensions imaged using 3D confocal microscopy \cite{VanBlaaderen1992,Dullens2003}. TPM droplets were dyed with a 1 mg/ml Cyanine 3 NHS Ester solution in acetone prepared using the same method described for RITC. Particles were transferred into solutions of refractive index matching 58:42 w/w TCE:TTL using the solvent transfer method described previously \cite{Liu2019}. The resulting suspension was centrifuged to form a pellet and a small volume of supernatant was removed and replaced with PERC. In this way, solutions of varying solvent density but identical TPM volume fraction could be prepared. Suspensions were sealed in a capillary and left to sediment overnight before imaging, focusing on the different diffusion-sedimentation equilibria created for different PERC volume fractions.

Note that all organic solvents contained 5\%wt OLOA 11000, a charge control agent and stabilizer which has been shown to keep TPM particles stable and hard-sphere-like in their interactions \cite{Yanagishima2021}.

\subsection{Imaging Methods}
All particles were sized by Scanning Electron Microscopy (SEM). Particles were dried on a silicon chip then sputter coated with platinum using a SC7620 sputter coated (Quorum Technologies, UK) before imaging with a JSM-G1010LV SEM (JEOL, Japan) using an acceleration voltage of 20kV. Images were analyzed using a circle finding algorithm within Matlab. At least 100 particles were imaged to obtain the mean diameter and polydispersity.

Confocal imaging was carried out on an Olympic 1X73 microscope equipped with a Thorlabs confocal 12kHz resonant point scanner, a 532nm laser, and standard FITC/Rhodamine filter cube. All measurements used a 60x Olympus Plan Achromatic oil immersion objective. Integrated particle brightness was calculated using a widely used particle identification algorithm \cite{Crocker1996}.

\section{Results and Discussion}

\subsection{TPM Droplet Synthesis}
\begin{figure}[H]
\includegraphics[width=15cm]{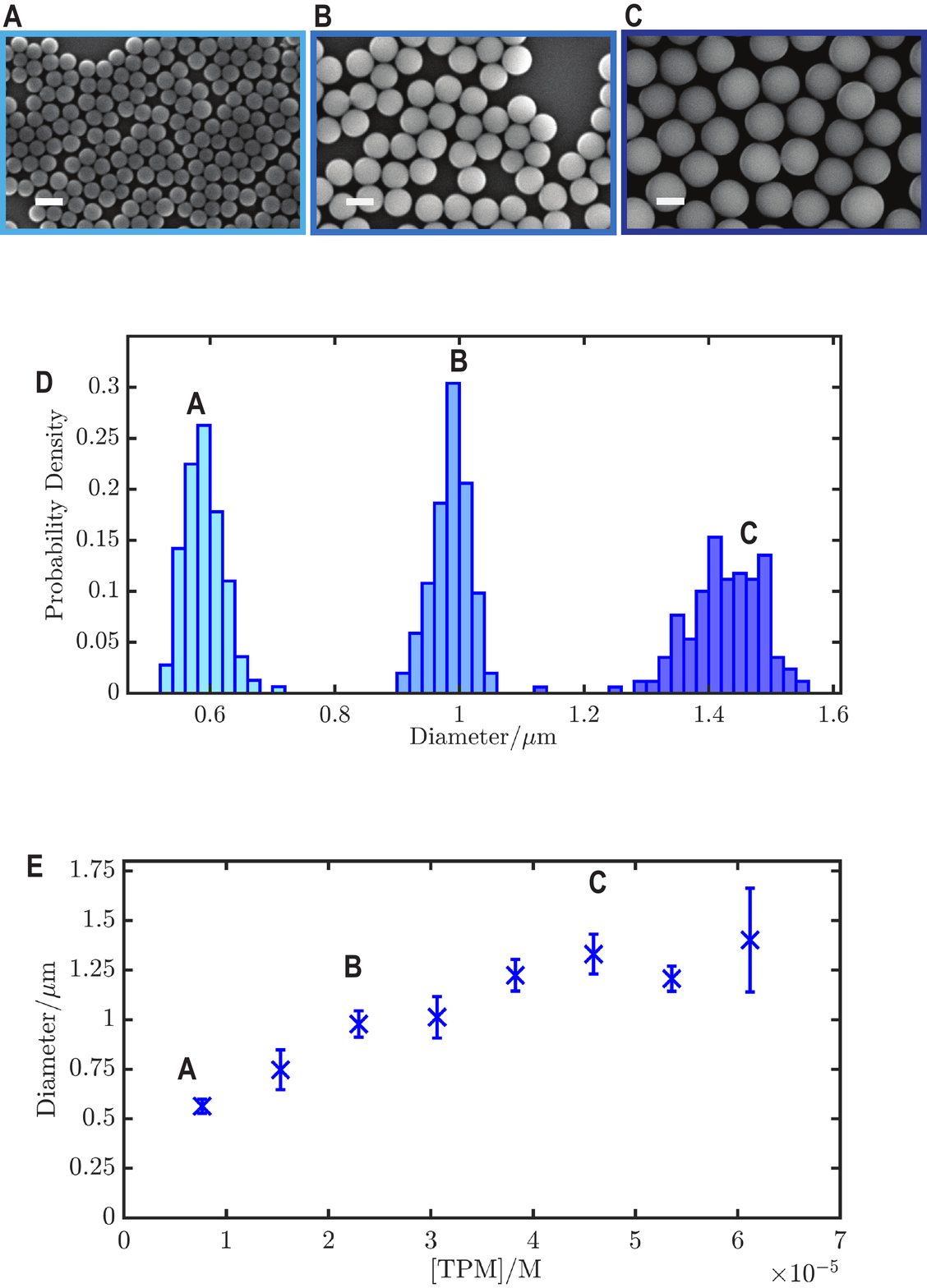}
\centering
\caption{SEM images of TPM particles synthesized from different initial concentrations of TPM: 0.77x$10^{-5}$M (A), 2.3 x$10^{-5}$M (B) and 3.83x$10^{-5}$M (C). Scale bars are 1 $\mu$m. (D) shows a histogram of particles sizes for A, B and C. (E) shows a plot of final diameter against concentration of TPM. Mean diameters shown are an average over at least four syntheses; error bars show the standard deviation of the means.}
\label{fig:qtpm}
\end{figure}

As this synthesis takes place in 'zero-flow' conditions, only two variables affect particle size: pH and concentration of TPM \cite{VanDerWel2017,Middleton2019}. Given a constant pH, we show that particle diameter can be precisely tuned by varying the concentration of TPM. Figures \ref{fig:qtpm}A-C show SEM images of TPM particles nucleated with increasing initial concentrations of TPM; there is a clear increase in particle size. No particles were nucleated below the concentrations shown. Importantly, even though no stirring is carried out, the polydispersity remains low in all cases, below 6\% for all batches; size distributions are given in Figure \ref{fig:qtpm}D.

The average particle diameter as a function of hTPM concentration is given in Figure \ref{fig:qtpm}E. Firstly, there is a broad consistency with the size of 0.5$\mu$m indicated in previous work \cite{Sacanna2011}. For low hTPM concentrations, there is a proportional increase in particle size with concentration. Note that the error bars indicate the standard deviation of mean diameters over three independently produced batches: this suggests that one can aim for a particle size of up to approximately 1.5 $\mu$m within a range of $\pm$ 0.1 $\mu$m. A small amount of flow is indeed introduced by addition of the base, but by using a pipette and pipette filler we found flow was negligible. This resulted in the same conditions for nucleation between batches and consequently an accurate synthesis of particles of a specific target size within an error of 100 nm. We do note that the last point deviates from this trend with significantly larger error bars. Many factors may contribute to this; for example, the flask becomes turbid significantly faster than at lower concentrations, increasing the likelihood that the minimal flow created by adding the base affects the nucleation.

The benefits become more apparent when we compare this to what happens using protocols with stirring. For example, Liu {\it et al.} and Van Der Wel {\it et al.} both synthesize particles at a monomer fraction of 20 mM in pH 10.7 conditions, but the resulting particle diameters are 1.3  $\mu$m and 0.6 $\mu$m respectively. In fact, Middleton {\it et al.} explicitly demonstrates that stirring rate has a non-monotonic impact on the final particle size. By increasing stirring speed from 500 to 900 min$^{-1}$ resulting particle diameter increases from 1.45 $\mu$m to 1.92 $\mu$m. However a higher stirring speed of 1100 min$^{-1}$ decreased particle size to 1.75 $\mu$m. Even with identical stirring equipment, identical flow fields are difficult to replicate between batches, and thus lead to discrepancies at the same concentration and pH. We propose that by eliminating the need for stirring by using pre-hydrolyzed TPM, which is already evenly dispersed throughout the reaction vessel, this method is more reproducible across reaction vessels. Since reactions can be done in anything from 0.5ml microtubes to large flasks, dozens of syntheses are trivially parallelized.

Droplet nucleation using pre-hydrolyzed TPM is also drastically faster than in stirred, one-pot syntheses: solutions turn turbid within 1 minute. This is likely a product of the formation mechanism. Though a variety of mechanisms have been proposed for particle formation in similar systems\cite{Whitehead2021}, there is a broad consensus that the rate limiting step in this process is the slow hydrolysis of the TPM under basic condition. Sacanna's method bypasses this step by using a pre-hydrolyzed TPM, leading to a rapidly developing turbidity \cite{Sacanna2011}. Importantly, to maximize control over nucleated droplet diameter, our data suggests that the best results are obtained when particle nucleation occurs after the residual flows from the mixing process have settled. In the parameter space explored in this work, this corresponds to low TPM concentrations, when the nucleation is comparatively slower. We also note that the ammonia concentration we use is significantly lower than previous work, again slowing down nucleation enough to see any residual flows dissipate. Though the elucidation of the exact mechanism is beyond the scope of this paper, the ease with which particles may be formed in the absence of flow suggest applications to in-situ particle formation in confined geometries given an arbitrary pH modification mechanism. 

\begin{figure}[H]
\includegraphics*[width=15cm]{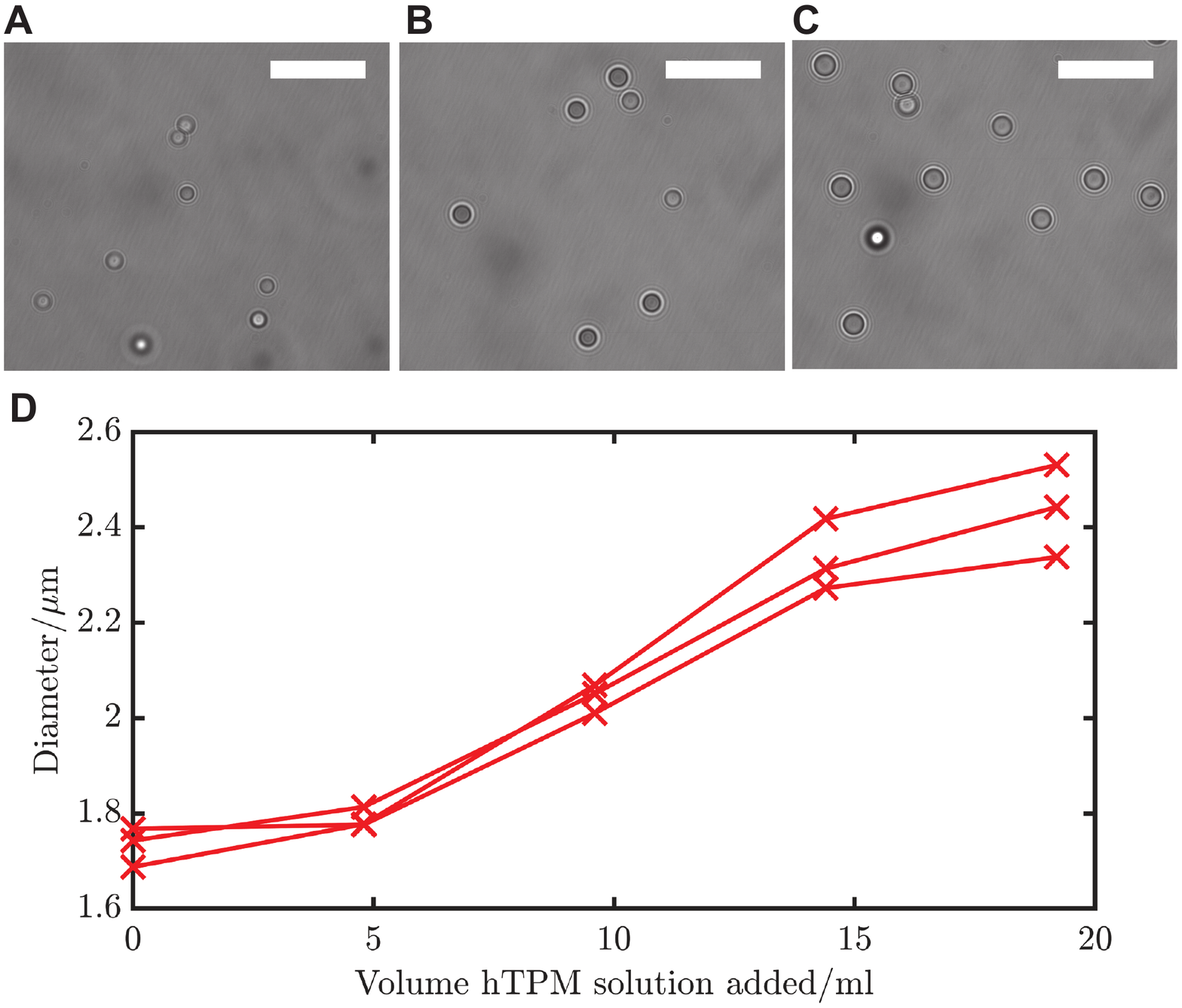}
\centering
\caption{(A)-(C) show optical microscope images taken during a droplet growth procedure. Scale bar for all images is 10 $\mu$m (D) shows a plot of particle diameter against volume of hTPM added for droplet growth experiments.}
\label{fig:growth}
\end{figure}

To produce particles larger than 1.5 $\mu$m, a droplet growth procedure was required, as in previous works \cite{Sacanna2011,VanDerWel2017}. Here, we document particle size growth with a set drip rate, with a focus on the reproducibility of the growth trajectory. Using a syringe pump to drip in pre-hydrolyzed TPM as described in the Methods, Figure \ref{fig:growth}A-C show optical microscopy images of droplets during the growth procedure. Note that there is no secondary nucleation or an increase in polydispersity. This is also shown in Figure \ref{fig:growth}D, showing particle diameter as a function of volume of hTPM added. An increase in particle diameter is seen, but polydispersity stays low, below 6$\%$. Slow addition of pre-hydrolyzed hTPM ensures the concentration of TPM is even throughout the flask, for even growth of all droplets, but too low for secondary nucleation. Importantly, the polydispersity of the resulting particles remains below the crystallization threshold \cite{Pusey1987-2}, making them suitable for studies of colloidal crystallization. Again, due to the faster incorporation of hTPM compared to TPM oil hydrolysing in-situ, growing particles up to 2.5 $\mu$m in diameter only takes 2 hours after initial nucleation. Looking at multiple growth trajectories, we find that independent growths lead to precise, quantitative control over diameter growth using this protocol: the three separate lines indicate independent droplet growths which follow the same path to within $\pm$0.1 $\rm{\mu}$m. We note that this droplet growth procedure can be performed on any of the initial droplet sizes nucleated under the conditions shown in Figure \ref{fig:growth}. Note that we have chosen a growth rate that suppresses secondary nucleation. If one was aiming for larger droplets in the same amount of time, it is also possible to raise the drip rate. This will induce secondary particle formation: however, this is not a problem, provided that the growth is stopped before the secondaries grow too much and cause issues for separation of the primary particles by centrifugation.

\subsection{Uniform Dyeing}
\begin{figure}[H]
\includegraphics*[width=15cm]{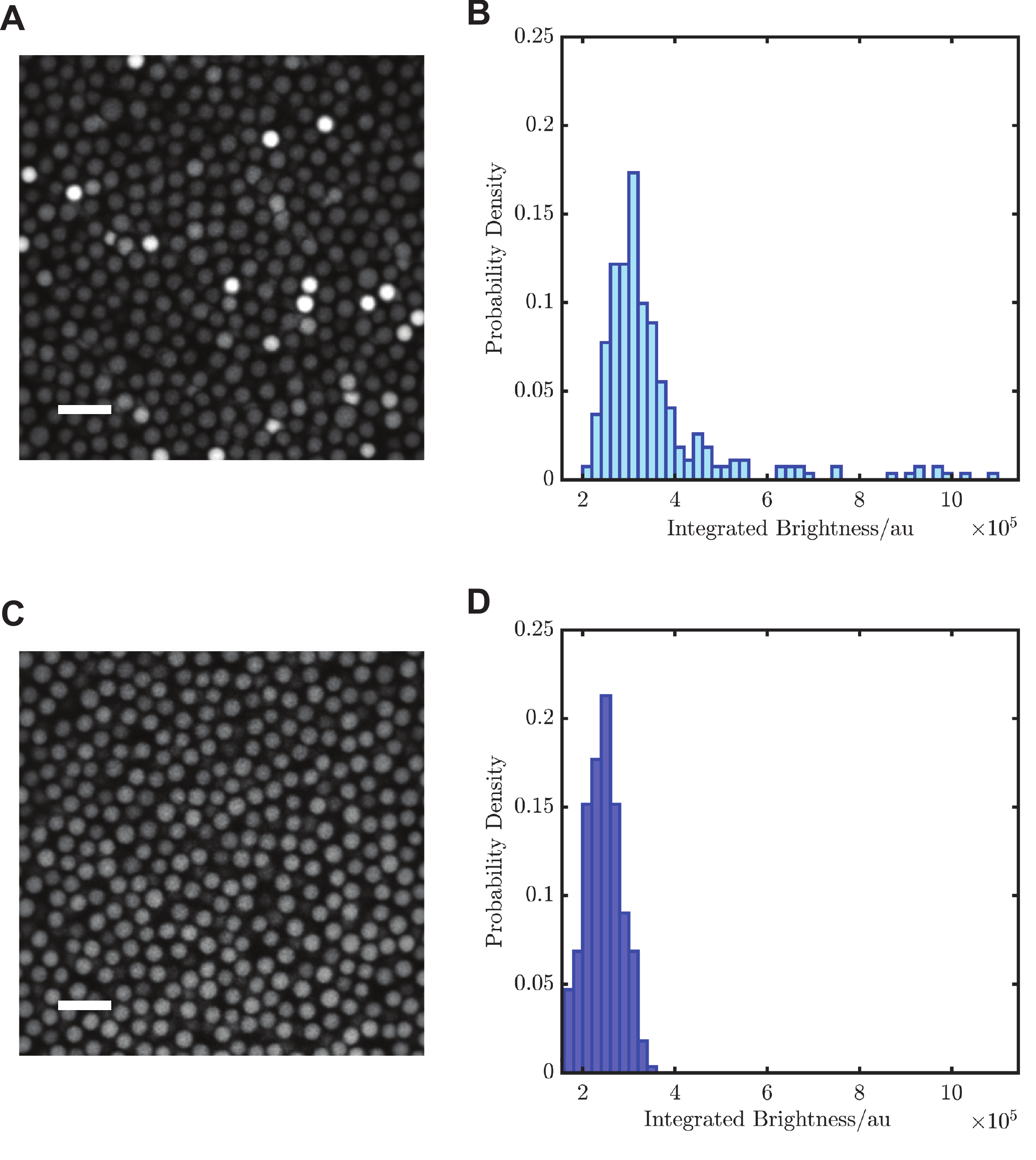}
\centering
\caption{2D fluorescent microscopy images of particles dyed with RITC using a DMSO (A) and acetone (C) dye solution. Scale bar is equal to 5 $\mu$m. (B) and (D) show histograms of integrated brightness for particles dyed with DMSO and acetone dyes respectively.}
\label{fig:dye}
\end{figure}

Most previous protocols to fluorescently dye TPM particles use a pre-labeled fluorophore dissolved in DMSO. Even with stirring, when the dye solution is added to TPM droplets, locally high concentrations of dye result where the dye solution is dripped into the reaction vessel, producing a small number of very bright particles which can interfere with the fluorescence profile of neighboring particles. We demonstrate this using a solution of RITC in DMSO, pre-conjugated with APS. Note the noticeably bright particles in Figure \ref{fig:dye}A and the elongated tail in the distribution of integrated particle brightness in Figure \ref{fig:dye}B. This can be avoided by changing the dye solvent from DMSO to acetone. Figure \ref{fig:dye}C is visibly more uniform; the distribution of integrated particle brightness in Figure \ref{fig:dye}D is also clearly monomodal, with no long tail.

To achieve more uniform transfer of a fluorophore, the organic solvent must disperse quickly and evenly throughout the aqueous reaction media. This ensures there are no points of high concentration of the organic solvent, and therefore the fluorophore. This would generate a local population of overly bright particles as some particles are exposed to a higher concentration of dye during the transfer than others. Similarly to how we use an organic solvent to disperse a fluorophore into an aqueous media, analogous work \cite{MoritzBeck-Broichsitter} uses organic solvents to disperse a polymer during nanoprecipitation experiments. By dissolving polymer in solvents of similar miscibility but a higher diffusion coefficient, Moritz-Broichsitter {\it et al} saw a more even dispersion of polymer in the aqueous phase. By selecting acetone, which has a higher diffusion coefficient then DMSO, we can exploit the same effect. Acetone disperses the dye more evenly prior to transfer to TPM droplets resulting in the entire batch of droplets being more evenly dyed.

\subsection{Tuning of Refractive Index and Density Matching}
Finally, we introduce a unique ternary solvent combination for preparing index matched suspensions with varying solvent densities. Previous work \cite{Liu2019,Yanagishima2021} describes how TPM microspheres may be approximately refractive index and density matched in a solution of TCE and TTL. Such suspensions may be imaged using confocal microscopy to yield the full internal 3D structure. However, the study of phenomena such as gelation and sedimentation often require independent adjustment of the density; in a binary solvent mixture, this would be at the sacrifice of index matching. As previously stated, PERC has the same refractive index but a much higher density than both TCE and TTL, and the crosslinked TPM microspheres. Therefore, addition of PERC to a refractive index matched suspension of TPM in TTL and TCE maintains the refractive index match while tuning the density mismatch.

\begin{figure}[H]
\includegraphics*[width=13.5cm]{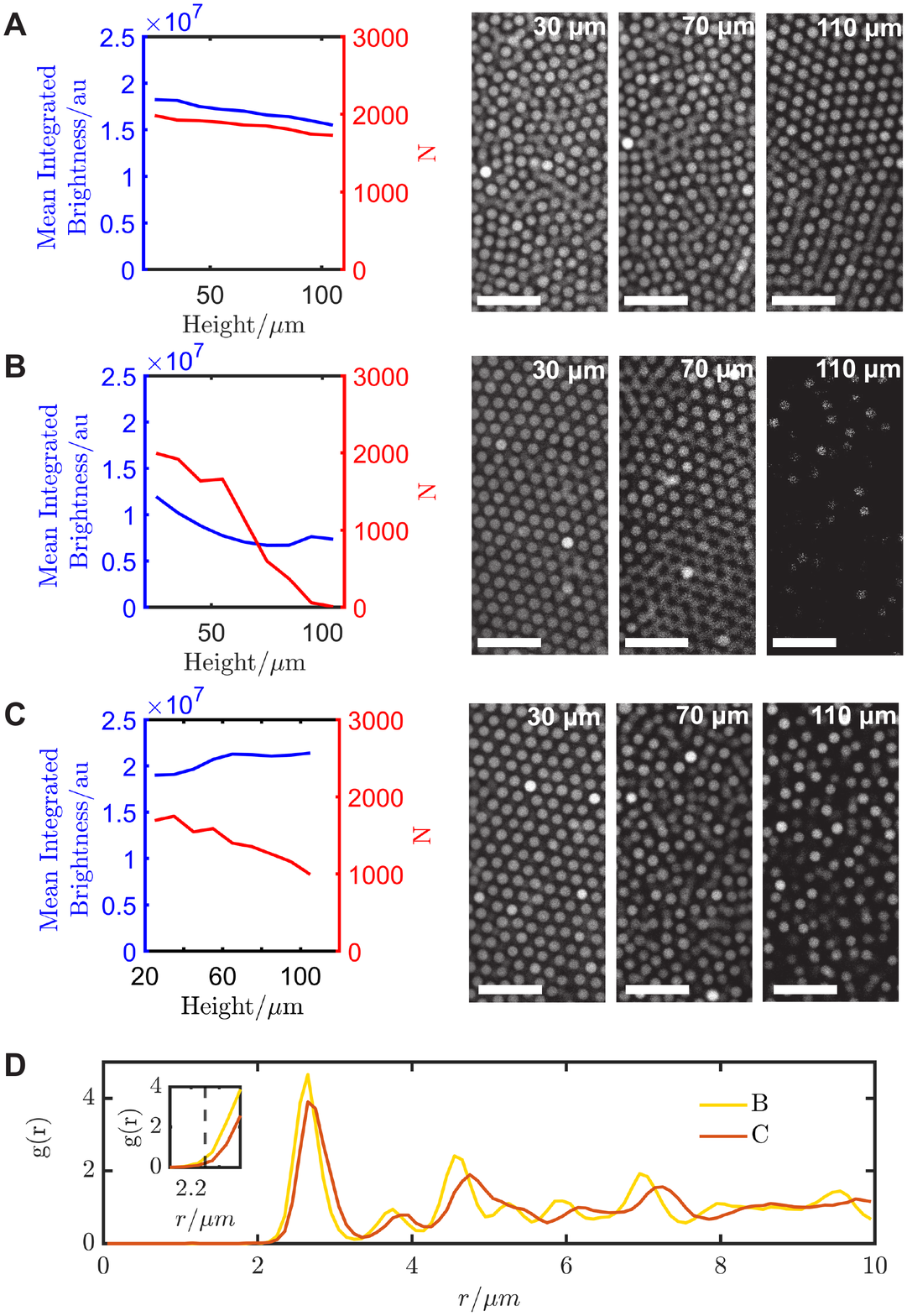}
\centering
\caption{Number of particles and mean integrated brightness per particle at different heights within a confocal z-stack through a partially crystalline sediment of TPM spheres in refractive index matching 42:58 TTL:TCE (A), in (nearly) density matching 85:15 w/w TCE:TTL (B) and (nearly) density matching 45:23:32 w/w PERC:TCE:TTL (C). Inserts show $xz$ profile at different heights. Scale bar in all cases is 10 $\mu$m. (D) is a 3D g(r) of samples B and C, inset shows the g(r) at small r. Dotted line shows the Diameter measure from SEM..}
\label{fig:confocal}
\end{figure}

Figure \ref{fig:confocal}A shows 3D confocal microscopy images of a suspension of dyed TPM particles with an additional non-fluorescent shell \cite{Liu2016} suspended in 42:58 w/w TCE:TTL, at a particle volume fraction of 30\%, left to sediment overnight in a capillary. The particles are 2.4 $\mu$m in diameter (see below) and are monodisperse. Two samples of this particle suspension in 42:58 w/w TCE:TTL are then taken; both are taken close to density matching by i) addition of TCE and ii) addition of PERC. We do not fully density match to ensure the formation of a relatively dense suspension at the base, to gauge imaging performance in dense suspensions. Figure \ref{fig:confocal}B shows the particles suspended in a solvent consisting of 85:15 w/w TCE:TTL. Note the decay in signal per particle as we go deeper into the suspension. This is also apparent in the image quality as scan depth increases. This emphatically shows that density matching cannot be achieved in a TCE:TTL binary system without sacrificing the refractive index matching. Figure \ref{fig:confocal}C, however, shows the results for the \emph{same} particle suspension with the same solvent density, only now using PERC with an overall composition 45:23:32 w/w PERC:TCE:TTL. The snapshots taken throughout the suspension show minimal apparent decay in brightness. The mean integrated brightness per particle as a function of height corroborate this. Note in particular the change in the number of particles detected, particularly at deep positions: this highlights the possibility that a significant number of particles may have been missed in Figure \ref{fig:confocal}B due to a poor signal-to-noise ratio.

Thus, we can independently tune the density mismatch relative to the particles while maintaining the refractive index matching, something not feasible for a binary system as demonstrated in Figure \ref{fig:confocal}B. We emphasize the ease with which one can tune the density mismatch: the suspension can simply be centrifuged to form a pellet and a volume of refractive index matching solvent removed and replaced with PERC. In this way a range of samples of the same refractive index matching but varying density mismatch may be prepared. We also note that no significant change in swelling is observed. Figure \ref{fig:confocal}D shows the g(r) for the close-to-density matched suspensions B and C. We see no clear shift in where the first peak rises (see inset) nor the position of the first peak, indicating minimal change in physical size.

\begin{figure*}[h]
\includegraphics*[width=12cm]{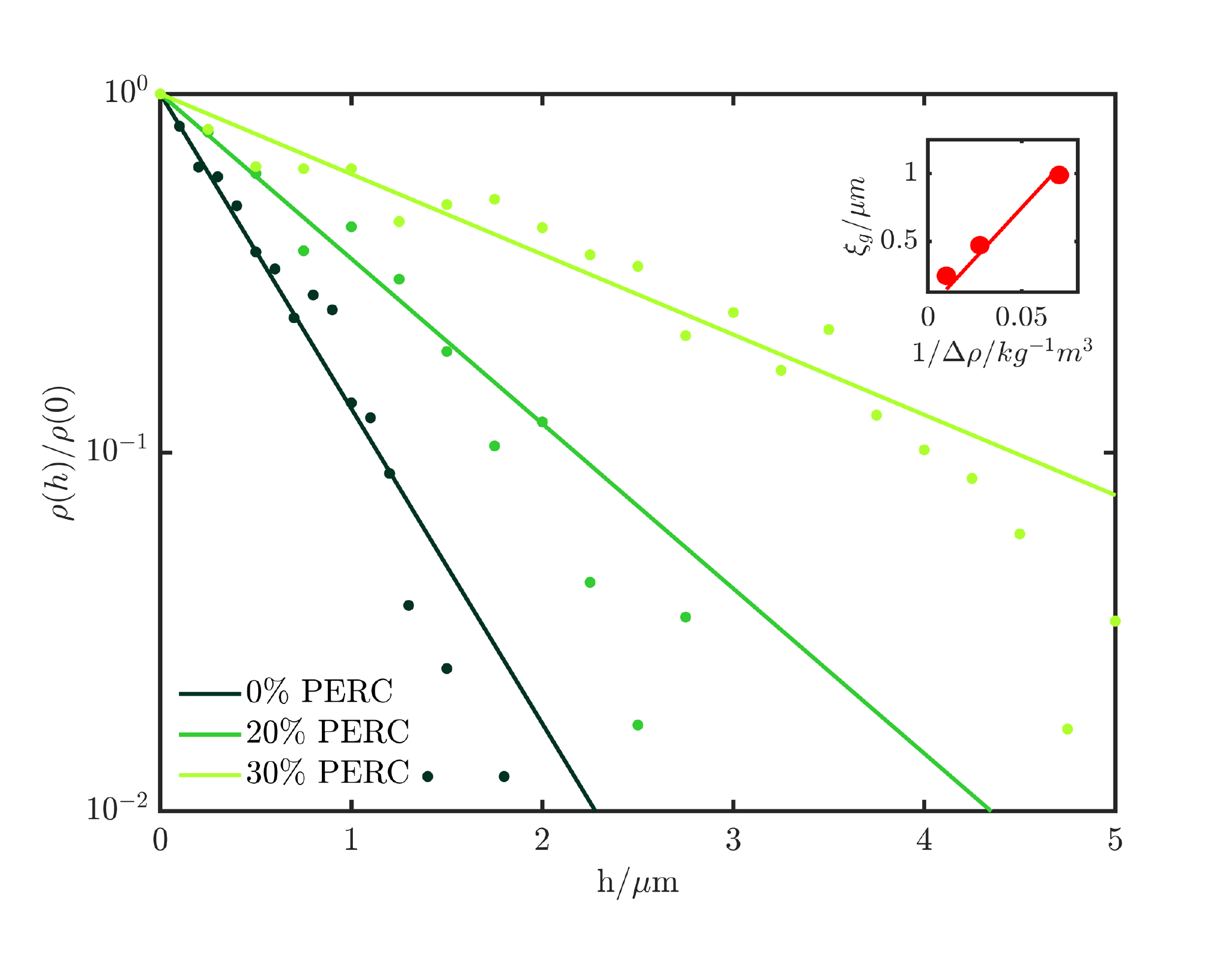}
\centering
\caption{Number density $\rho(h)$ against height for particles suspended in a ternary mixture of TCE:TTL:PERC with varying volume fractions of PERC. The solid line is an exponential fit. (inset) Gravitational length vs. inverse of the density difference between the particle and the solvent. The gradient may be used to extract a particle size.}
\label{fig:sedimentation}
\end{figure*}

The use of PERC makes it simple to realize a series of solutions with varying density mismatch. For a sedimentation-diffusion equilibria of colloidal spheres, this mismatch is characterized by the characteristic length scale of the decay in number density profile $\rho(h)$, also known as the gravitational length $\xi_{g}$. $\rho(h)$ in the dilute limit is described by the barometric formula, $\rho_{h}=\rho_{0}\exp(-\frac{h}{\xi_{g}})$. Figure \ref{fig:sedimentation} shows this profile for varying concentrations of PERC on a semilog plot, accompanied by fits to a simple exponential decay. The correspondence is very clear. Note that $\xi_g$ is expressed as
\begin{equation}
    \xi_g = \frac{k_{\rm B}T}{\left(\Delta\rho\right)\left(\frac{1}{6}\pi{d^3}\right)g} ,
\end{equation}
where $k_BT$ is the thermal energy, $\Delta\rho$ is the density difference between the particle and the solvent, $d$ is the particle diameter and $g$ is acceleration due to gravity. Thus, a reduced density mismatch leads to a longer $\xi_g$; this is also reflected in the data. In fact, with an easily prepared series of densities and barometric profiles, we may plot $\xi_g$ against $1/\Delta\rho$ to estimate the swelled particle size, assuming no drastic changes due to the addition of PERC (inset of Figure \ref{fig:sedimentation}). We find this gives a diameter of 2.4 $\mu$m of the particles in solution. This is consistent with the value measured from SEM, 2.3 $\mu$m; compared to dried particles in a vacuum, particles are expected to be swelled \cite{Liu2019}.

The series of solvents used above span P\'{e}clet numbers ranging from approximately 0.6 to 2.4. They can be made as arbitrarily low by adding more PERC, or higher by preparing larger particles. Importantly, for particles of a few micrometers, it is evidently simple to probe the regime where the P\'{e}clet number is approximately 1, where thermal and advective effects may compete. This illustrates how a ternary system of TCE, TTL and PERC is ideal for studying colloidal suspensions with a controlled P\'{e}clet number using confocal microscopy without any compromise in the refractive index matching.

\clearpage
\section{Conclusion}
We have characterized and optimized a simple and reproducible 'zero-flow' protocol for TPM synthesis \cite{Sacanna2010,Sacanna2011}, demonstrating that single-step nucleation can realize precise control over particle size. The method is free of dependencies around reactant flow, and highly reproducible between independent attempts. Droplets may be grown further by adding more pre-hydrolyzed TPM at a constant rate, which is consistent with other works \cite{Sacanna2011}. We also present improvements to existing methods of dyeing TPM which improve the distribution of dye over the entire batch of particles. Finally, we introduce a ternary organic solvent mixture which allows independent control of density mismatch within a refractive index matched sample, demonstrated by direct examination of sedimentation-diffusion equilibria. The ease of synthesis, imaging using confocal microscopy and tuning of sedimentation length make this an excellent model system for investigating a wide range of condensed phase behaviour, particularly when particle size distribution \cite{Pusey1987-2,Schope2007,Royall2013} or gravity have a significant role to play.

\section{Acknowledgments}
BvdM acknowledges funding from the Netherlands Organization for Scientific Research (NWO) through a Rubicon grant (NWO-Rubicon Grant No. 019.191EN.011). RPAD acknowledges the European Research Council (ERC Consolidator Grant No. 724834 OMCIDC) for financial support. TY acknowledges a Kyoto University Internal Grant for Young Scientists (Start-up), a Toyota Riken Scholar Grant from the Toyota Physical and Chemical Research Institute, and the Japan Science and Technology Agency (JST) CREST Program Grant Number JPMJCR2095.

\bibliography{references}

\end{document}